\documentclass[11pt]{article}
\pdfoutput=1

\usepackage[utf8]{inputenc}
\usepackage{amsmath,amssymb,amsthm}
\usepackage{array}
\usepackage{authblk}
\usepackage[english]{babel}
\usepackage{bm}
\usepackage{booktabs}
\usepackage{caption}
\usepackage{courier}
\usepackage{csquotes}
\usepackage{enumitem}
\usepackage[margin=1.0in]{geometry}
\usepackage{graphicx}
\usepackage{hyperref}
\usepackage{mathtools}
\usepackage[square,numbers]{natbib}
\usepackage{subfiles}

\graphicspath{ {images/} }

\hypersetup{
    colorlinks=true,
    citecolor=blue,
    linkcolor=black,
}

\title{\vspace{-1em}Association of genomic subtypes of lower-grade gliomas with shape features automatically extracted by a deep learning algorithm\\\vspace{0.2em}}
\date{\vspace{-2em}}

\author[1]{Mateusz Buda}
\author[1]{Ashirbani Saha}
\author[1, 2]{Maciej A. Mazurowski}

\affil[ ]{\footnotesize }
\affil[1]{\footnotesize Department of Radiology, Duke University, Durham, NC}
\affil[2]{\footnotesize Department of Electrical and Computer Engineering, Duke University, Durham, NC}

\begin{document}

\maketitle

\begin{abstract}

Recent analysis identified distinct genomic subtypes of lower-grade glioma tumors which are associated with shape features.
In this study, we propose a fully automatic way to quantify tumor imaging characteristics using deep learning-based segmentation and test whether these characteristics are predictive of tumor genomic subtypes.

We used preoperative imaging and genomic data of 110 patients from 5 institutions with lower-grade gliomas from The Cancer Genome Atlas.
Based on automatic deep learning segmentations, we extracted three features which quantify two-dimensional and three-dimensional characteristics of the tumors.
Genomic data for the analyzed cohort of patients consisted of previously identified genomic clusters based on IDH mutation and 1p/19q co-deletion, DNA methylation, gene expression, DNA copy number, and microRNA expression.
To analyze the relationship between the imaging features and genomic clusters, we conducted the Fisher exact test for 10 hypotheses for each pair of imaging feature and genomic subtype.
To account for multiple hypothesis testing, we applied a Bonferroni correction.
P-values lower than 0.005 were considered statistically significant.

We found the strongest association between RNASeq clusters and the bounding ellipsoid volume ratio ($p<0.0002$) and between RNASeq clusters and margin fluctuation ($p<0.005$).
In addition, we identified associations between bounding ellipsoid volume ratio and all tested molecular subtypes ($p<0.02$) as well as between angular standard deviation and RNASeq cluster ($p<0.02$).
In terms of automatic tumor segmentation that was used to generate the quantitative image characteristics, our deep learning algorithm achieved a mean Dice coefficient of 82\% which is comparable to human performance.

\vspace{1em}
\noindent
\textbf{Keywords:} deep learning; brain segmentation; radiogenomics; MRI; LGG

\end{abstract}

\section{Introduction}
\label{sec:introduction}

Lower-grade gliomas (LGG) are a group of WHO grade II and grade III brain tumors.
As opposed to grade I which are often curable by surgical resection, grade II and III are infiltrative and tend to recur and evolve to higher-grade lesion.
Predicting patient outcomes based on histopathological data for these tumors is inaccurate and suffers from inter-observer variability~\cite{cancer2015comprehensive}.
One of the promising methods that might address this issue is defining subtypes of LGG through clustering of patients based on DNA methylation, gene expression, DNA copy number, and microRNA expression~\cite{cancer2015comprehensive}.
It was shown that the clusters identified in such way are to a large extent in agreement with another basic molecular subtype based on IDH (IDH1 and IDH2) mutation and 1p/19q co-deletion~\cite{cancer2015comprehensive, zhang2016genomic}.
Patients with tumors from different molecular groups substantially differ in terms of typical course of the disease and overall survival~\cite{eckel2015glioma}.

A new research direction in cancer, called radiogenomics, aims at investigating the relationship between tumor genomic characteristics and medical imaging~\cite{mazurowski2015radiogenomics}.
Imaging can provide important information before surgery or in cases when resection is not possible.
Very recent studies in this area have discovered an association of tumor shape features extracted from MRI with its genomic subtypes~\cite{mazurowski2017radiogenomic, mazurowski2017radiogenomics}.
However, the first step when extracting tumor features was the manual segmentation of MRI. Such annotation is costly, time consuming and results in annotations with high inter-rater variance~\cite{menze2014multimodal}.

Deep learning is a new field of machine learning that is recently revolutionizing the automated analysis of images~\cite{goodfellow2015deep, ioannidou2017deep}.
There are many examples of successful applications of deep learning in medical imaging~\cite{ronneberger2015u, cciccek20163d, milletari2016v, chen2018voxresnet, mazurowski2019deep} and more specifically in brain MRI segmentation~\cite{havaei2016deep}.
In recent years, progress in deep learning for automatic brain segmentation matured to a level that achieves performance of a skilled radiologist~\cite{havaei2017brain}.
Most of these efforts are focused on glioblastoma rather than LGG.
Development of models that yield high quality segmentation of LGG in brain MRI would potentially allow for automatization of the process of tumor genomic subtype identification through imaging that is fast, inexpensive, and free of inter-reader variability.

In this study, we combine the field of deep learning and radiogenomic and propose a fully automatic algorithm for quantification of tumor shape and test whether the assessed shape features are prognostic of tumor molecular subtypes.
Developing imaging-biomarkers that could inform of tumor genomics would provide the information to clinicians sooner in a non-invasive way and in some cases could allow for better stratification of tumors where resection is not performed.
In this study, we show promise for eventually developing such imaging-based biomarkers.

The reminder of this paper is organized as follows.
Section~\ref{sec:dataset} describes data used in our study whereas section~\ref{sec:methods} describes segmentation model, features used for tumor quantification, and statistical methods.
Then, in section~\ref{sec:results} we show results for the segmentation algorithm and prediction of genomic subtypes.
In section~\ref{sec:discussion} we discuss our findings. Finally, sections~\ref{sec:limitations} and~\ref{sec:conclusions} are devoted to limitations and conclusions of the study, respectively.

\section{Dataset}
\label{sec:dataset}

\subsection{Patient population}
The data used in this study was obtained from The Cancer Genome Atlas (TCGA) and The Cancer Imaging Archive (TCIA).
We identified 120 patients from TCGA lower-grade glioma collection\footnote{\url{https://cancergenome.nih.gov/cancersselected/lowergradeglioma}} who had preoperative imaging data available, containing at least a fluid-attenuated inversion recovery (FLAIR) sequence.
Ten patients had to be excluded since they did not have genomic cluster information available.
The final group of 110 patients was from the following 5 institutions: Thomas Jefferson University (TCGA-CS, 16 patients), Henry Ford Hospital (TCGA-DU, 45 patients), UNC (TCGA-EZ, 1 patient), Case Western (TCGA-FG, 14 patients), Case Western – St. Joseph’s (TCGA-HT, 34 patients) from TCGA LGG collection.
The complete list of patients used in this study is included in Online Resource 1.
The entire set of 110 patients was split into 22 non-overlapping subsets of 5 patients each.
This was done for evaluation with cross-validation.

\subsection{Imaging data}
Imaging data was obtained from The Cancer Imaging Archive\footnote{\url{https://wiki.cancerimagingarchive.net/display/Public/TCGA-LGG}} which contains the images corresponding to the TCGA patients and is sponsored by the National Cancer Institute.
We used all modalities when available and only FLAIR in case any other modality was missing.
There were 101 patients with all sequences available, 9 patients with missing post-contrast sequence, and 6 with missing pre-contrast sequence.
The complete list of available sequences for each patient is included in Online Resource 1.
The number of slices varied among patients from 20 to 88.
In order to capture the original pattern of tumor growth, we only analyzed preoperative data.
The assessment of tumor shape was based on FLAIR abnormality since enhancing tumor in LGG is rare.

A researcher in our laboratory, who was a medical school graduate with experience in neuroradiology imaging, manually annotated FLAIR images by drawing an outline of the FLAIR abnormality on each slice to form training data for the automatic segmentation algorithm.
We used software developed in our laboratory for this purpose.
A board eligible radiologist verified all annotations and modified those that were identified as incorrect.
Dataset of registered images together with manual segmentation masks for each case used in our study is released and made publicly available at the following link: \url{https://kaggle.com/mateuszbuda/lgg-mri-segmentation}.

\subsection{Genomic data}
Genomic data used in this study consisted of DNA methylation, gene expression, DNA copy number, and microRNA expression, as well as IDH mutation 1p/19q co-deletion measurement.
Specifically, in our analysis we consider six previously identified molecular classifications of LGG that are known to be correlated with some tumor shape features~\cite{mazurowski2017radiogenomics}:
\begin{enumerate}[noitemsep,topsep=0em]
    \item Molecular subtype based on IDH mutation and 1p/19q co-deletion (three subtypes: IDH mutation-1p/19q co-deletion, IDH mutation-no 1p/19q co-deletion, IDH wild type)
    \item RNASeq clusters (4 clusters: R1-R4)
    \item DNA methylation clusters (5 clusters: M1-M5)
    \item DNA copy number clusters (3 clusters: C1-C3)
    \item microRNA expression clusters (4 clusters: mi1-mi4)
    \item Cluster of clusters (3 clusters: coc1-coc3)
\end{enumerate}

\section{Methods}
\label{sec:methods}

\subsection{Automatic segmentation}
\begin{figure}[!ht]
    \centering
    \includegraphics[width=\linewidth]{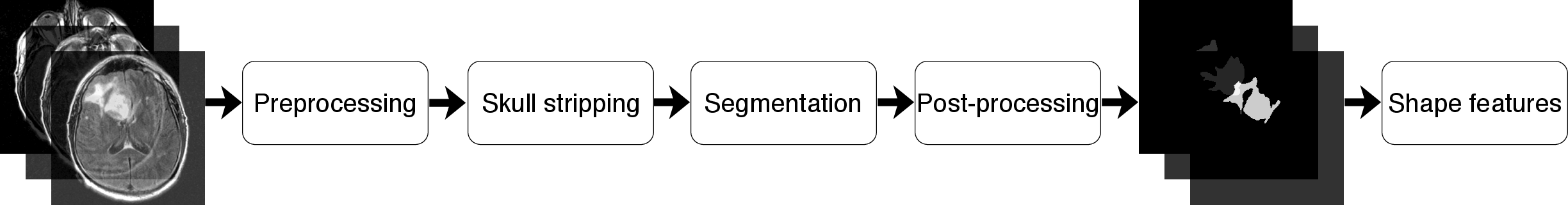}
    \caption{A schema showing data processing steps of our system for molecular subtype inference from a sequence of brain MRI}
    \label{fig:fig1}
\end{figure}

Figure~\ref{fig:fig1} shows the overview of the segmentation algorithm.
The following phases comprise the fully automatic algorithm for obtaining the segmentation mask: image preprocessing, segmentation, and post-processing.
Then, once the segmentation masks are generated, we extracted shape features that were identified as predictive of molecular subtypes.
The following sections provide details on each of the steps.
Source code of the algorithm described in this section is also available at the following link: \url{https://github.com/mateuszbuda/brain-segmentation}.

\subsubsection{Preprocessing}
Images varied significantly between patients in terms of size.
The preprocessing of the image sequences consisted of the following steps:
\begin{itemize}[noitemsep,topsep=0em]
    \item Scaling of the images to the common frame of reference.
    \item Removal of the skull to focus the analysis on the brain region (a.k.a., skull stripping).
    \item Adaptive window and level adjustment based on the image histogram to normalize intensities of tissues between cases.
    \item Z-score normalization of the entire data set.
\end{itemize}
The details of all the pre-processing steps are included in the Online Resource 2.

\subsubsection{Segmentation}
\begin{figure}[!ht]
    \centering
    \includegraphics[width=\linewidth]{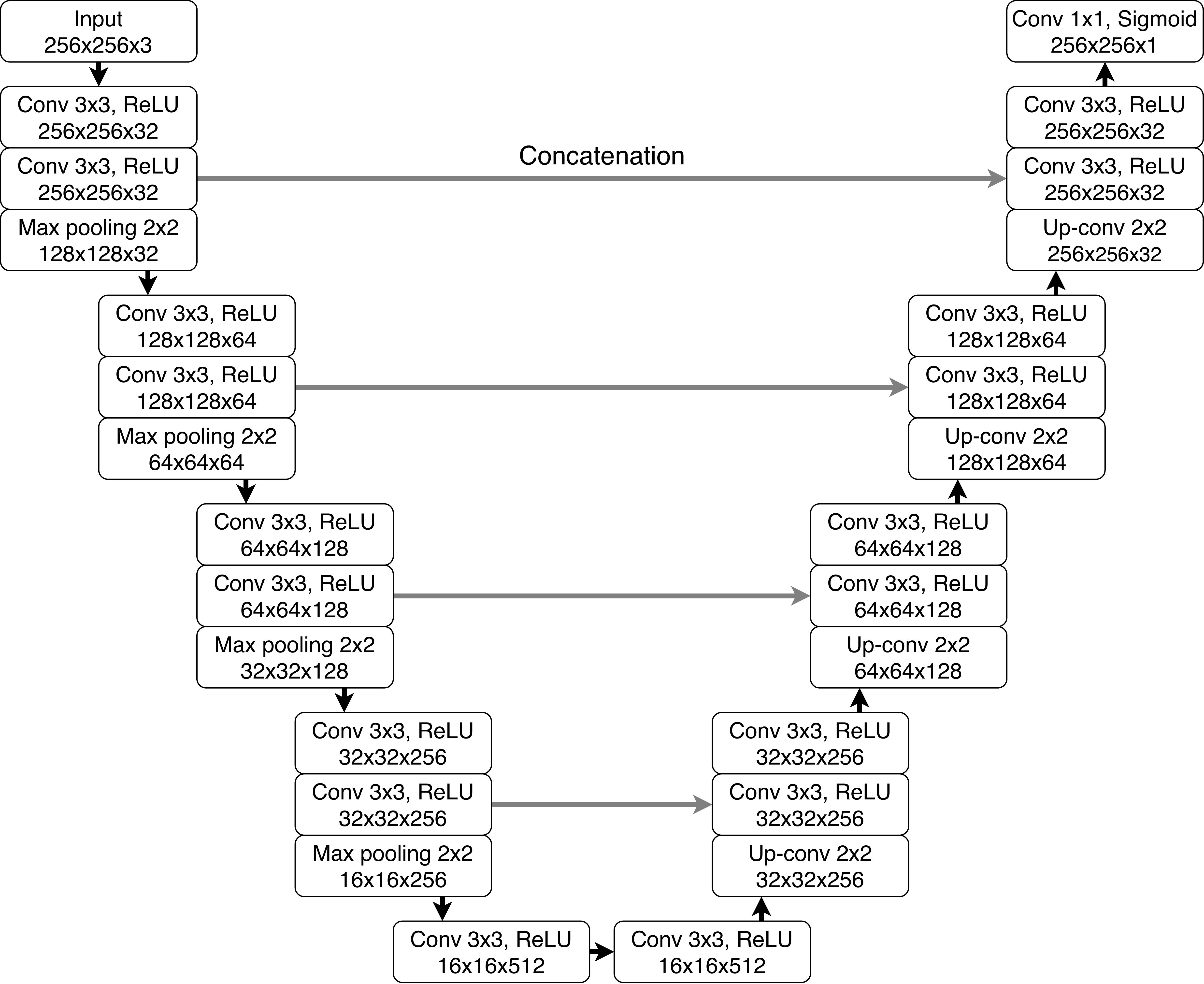}
    \caption{U-Net architecture used for skull stripping and segmentation. Below each layer specification we provide dimensionality of a single example that this layer outputs}
    \label{fig:fig2}
\end{figure}

The main segmentation step was performed using a fully convolutional neural network with the U-Net architecture~\cite{ronneberger2015u} shown in Figure~\ref{fig:fig2}.
It comprises four levels of blocks containing two convolutional layers with ReLU activation function and one max pooling layer in the encoding part and up-convolutional layers instead in the decoding part~\cite{glorot2011deep, lecun1995convolutional, lecun1998gradient, krizhevsky2012imagenet, noh2015learning}.
Consistent with the U-Net architecture, from the encoding layers we use skip connections to the corresponding layers in the decoding part.
They provide a shortcut for gradient flow in shallow layers during the training phase~\cite{drozdzal2016importance}.

Manual segmentation served as a ground truth for training a model for automatic segmentation.
We trained two networks, one for cases with three sequences available (pre-contrast, FLAIR, and post-contrast) and the other that used only FLAIR.
For the second network, instead of missing sequences we used neighboring FLAIR slices from both sides of a slice of interest as additional channels.
Since in this scenario the two sequences, which occupied channel 1 and channel 3 of the input are not available, we filled these channels with neighboring tumor slices to provide additional information to the network.

The number of slices containing tumor was considerably lower than those with only background class present.
Therefore, to account for this fact, we applied oversampling with data augmentation that was proved to help in training convolutional neural networks~\cite{buda2018systematic}.
We did it by having three instances of each tumor slice in our training set.
For one oversampled slice we applied random rotation by 5 to 15 degrees and for the other slice we applied random scale by 4\% to 8\%.
To further reduce the imbalance between tumor and non-tumor pixels, we discarded empty slices that did not contain any brain or other tissue after applying skull stripping.
This step has been undertaken since training a fully convolutional neural network with images that do not contain any positive voxels can be highly detrimental.
Please note that a significant majority of voxels in the abnormal slices are still normal and therefore sufficient negative data is available for training.

\subsubsection{Post-processing}
To further improve the accuracy, we implemented a post-processing algorithm that removes false positives.
Specifically, we extracted all tumor volumes using connected components algorithm on a three-dimensional segmentation mask for each patient.
We did it using 6-connected pixels in three dimensions, i.e. neighboring pixels are defined as being connected along primary axes.
Eventually, we included in the final segmentation mask only the pixels comprising the largest connected tumor volume.
This post-processing strategy benefits extraction of shape features (described in the following section) since they are sensitive to isolated false positive pixel segmentations.

\subsubsection{Extraction of shape features}
We consider three shape features of a segmented tumor that were identified as important in the context of lower grade glioma radiogenomic~\cite{mazurowski2017radiogenomics}:

\textit{Angular standard deviation (ASD)} is the average of the radial distance standard deviations from the centroid of the mass across ten equiangular bins in one slice, as described in~\cite{georgiou2007multi}.
Before calculating the value of this feature, we normalize radial distances to have mean equal one.
Angular standard deviation of a tumor shape is a quantitative measure of variation in the tumor margin within relatively small parts of the tumor.
It also captures non-circularity of the tumor, i.e. low value indicates circle like shape.

\textit{Bounding ellipsoid volume ratio (BEVR)} is the ratio between the volume of segmented FLAIR abnormality and its minimum bounding ellipsoid.
This feature captures the irregularity of the tumor in three dimensions.
If the tumor fits well into its bounding ellipsoid (high value of BEVR), it is considered more regular while if more space in the bounding ellipsoid is unfilled, the shape is considered irregular.

\textit{Margin fluctuation (MF)} is computed as follows.
First, we find the centroid of the tumor and distances from it to all pixels on the tumor boundary in one slice.
Then, we apply averaging filter of length equal to 10\% of the tumor perimeter measured in the number of pixels.
Margin fluctuation is the standard deviation of the difference between values before and after smoothing, i.e. applying averaging filter.
Similarly as in ASD, radial distances are normalized to have a mean of one.
This is done in order to remove the impact of tumor size on the value of this feature.
Margin fluctuation is a two-dimensional feature that quantifies the amount of high frequency changes, i.e. smoothness of the tumor boundary and was previously used for analysis of spiculation in breast tumors~\cite{giger1994computerized, pohlman1996quantitative}.

\subsection{Statistical analysis}
Our hypothesis was that fully automatically-assessed shape features are predictive of tumor molecular subtypes.
Since we considered 6 definitions of molecular subtypes based on genomic assays and multiple imaging features, we focused our analysis on the relationships between imaging and genomics that were found significant (with manual tumor segmentation) in a previous study~\cite{mazurowski2017radiogenomics}.
Specifically, those were the following relationships: bounding ellipsoid volume ratio with RNASeq, miRNA, CNC, and COC, the relationship of Margin fluctuation with RNASeq, and the relationship of angular standard deviation with IDH/1p19q, RNASeq, Methylation, CNC, and COC resulting in 10 specific hypotheses. To assess statistical significance of these associations, we conducted the Fisher exact test (fisher.test function in R) for each of 10 combinations of imaging and genomics.
For the purpose of this test, we turned each continuous imaging variable value into a number from 1 to 4 based on which quartile of the feature value it fell into.
For each of the imaging and genomic feature combinations, we used only the cases that had both: imaging and genomic subtype data available.

We conducted a total of 10 statistical tests for each pair of imaging feature and genomic subtype for our primary hypothesis.
To account for multiple hypothesis testing, we applied a Bonferroni correction.
P-values lower than 0.005 (0.05/10) were considered statistically significant for our primary radiogenomics hypotheses. 

Additionally, we evaluated performance of the deep learning-based segmentation itself.
We used Dice similarity coefficient~\cite{dice1945measures} as the evaluation metric which measures the overlap between the segmentation provided by the algorithm and the manually-annotated gold standard.

In the evaluation process, we used cross-validation.
Specifically, we divided our entire dataset into 22 subsets, each containing exactly 5 cases.
The model training was conducted on the training subsets and then the model was applied to the test cases.
This was repeated 22 times until each subset served once as the test set.
The cases then were pooled for the analysis as described above.

The number of cases included in the training and test sets (which determines the number of folds) is a trade-off between computational cost of training multiple models and having more data to train each of them.
The two extremes of this approach are leave-one-out strategy which results in one-case folds and the other is 50\% split which gives 2 folds.
We found folds of 5 patients to be a good balance between a training set size and computational cost.

\section{Results}
\label{sec:results}

The patients' characteristics are shown in Table~\ref{tab:tab1}.
The average patient age was 47.
Fifty six of the patients were women, fifty three were men, and the gender of one was unknown.
The tumors' characteristics are provided in Table~\ref{tab:tab2}.
There were three histological types of tumors: oligodendroglioma~(47), astrocytoma~(33), oligoastrocytoma~(29), and one unknown.
Grade of the tumors in our data included 51 cases of grade II, 58 of grade III, and grade of one tumor was unknown.

\begin{table}[!ht]
    \centering
    \begin{tabular}{l l} \toprule
    Characteristic & Patients (N=110) \\ \midrule
    Age (years) & \\
    ~~Median & 47 \\
    ~~Range & 20-75 \\ \midrule
    Gender & \\
    ~~Female & 56 \\
    ~~Male & 32 \\
    ~~Not available & 1 \\ \bottomrule
    \end{tabular}
    \caption{Patient characteristic. Age for one patient was missing and was ignored in the calculation.}
    \label{tab:tab1}
\end{table}

\begin{table}[!hb]
    \centering
    \begin{tabular}{l l} \toprule
    Characteristic & Cases (N=110) \\ \midrule
    Histologic type and grade & \\
    Astrocytoma & \\
    ~~Grade II & 8 \\
    ~~Grade III & 25 \\
    Oligoastrocytoma & \\
    ~~Grade II & 14 \\
    ~~Grade III & 21 \\
    Oligodendroglioma & \\
    ~~Grade II & 29 \\
    ~~Grade III & 18 \\
    ~~Not available & 1 \\ \midrule
    IDH-1p/19q subtype & \\
    ~~IDH mutation, 1p/19q co-deletion & 26 \\
    ~~IDH mutation, no 1p/19q co-deletion & 56 \\
    ~~IDH wild type & 25 \\
    ~~Not available & 3 \\ \bottomrule
    \end{tabular}
    \caption{Tumor characteristic.}
    \label{tab:tab2}
\end{table}

\begin{figure}[!h]
    \centering
    \includegraphics[width=0.55\linewidth]{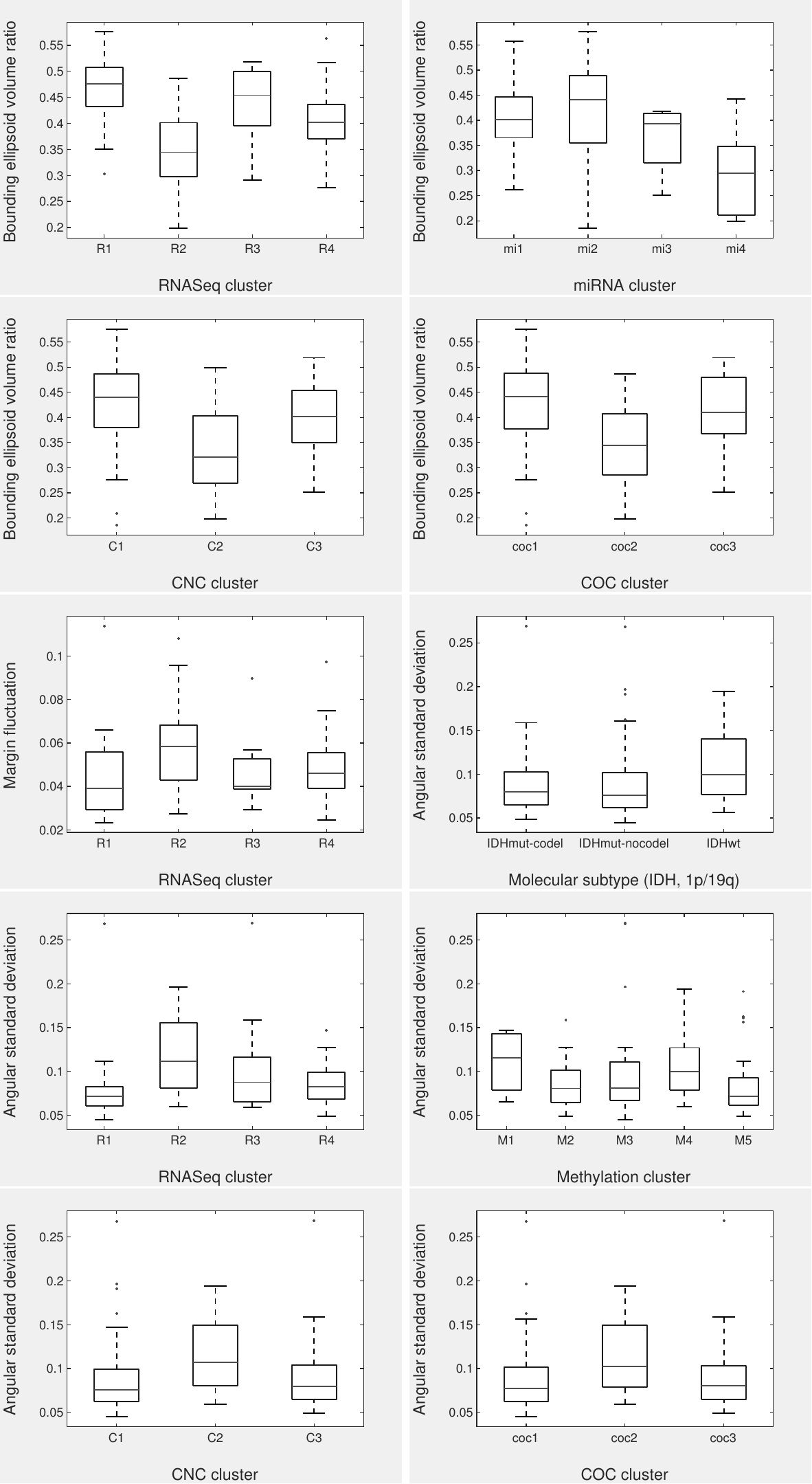}
    \caption{\textit{Box and whisker} plots demonstrating the relationship between tested genomic clusters and imaging features that quantify tumor shape}
    \label{fig:fig3}
\end{figure}

The results of the radiogenomic analysis are shown in Figure~\ref{fig:fig3}.
We confirmed our primary hypothesis for two pairs of imaging features and genomic subtype.
We found the strongest association between RNASeq cluster and the bounding ellipsoid volume ratio ($p<0.0002$) along with margin fluctuation ($p<0.005$).
In addition, we identified considerable correlations for the bounding ellipsoid volume ratio and all tested molecular subtypes ($p<0.02$) as well as for angular standard deviation and RNASeq cluster ($p<0.02$).

In Table~\ref{tab:tab3} we provide ROC AUC scores for the task of discriminating each RNASeq cluster from all other subtypes based on shape features extracted from segmentation masks obtained with the deep learning algorithm and compare them to manual segmentations.
We selected RNASeq cluster since it showed the strongest association with shape features.
In addition, we included ROC AUC based on two demographic variables, i.e. age and gender.
The results show that deep learning was able to provide tumor segmentations of a quality that allowed for extraction of shape features that match manual segmentations.
Specifically, cluster R2 was distinguished from all other clusters based on inversed bounding ellipsoid volume ratio with AUC of 0.80 and 0.78 for deep learning-based and manual segmentations, respectively.
In terms of angular standard deviation, AUC for deep learning was 0.73 and for manual segmentations was 0.72.
The predictive value of demographic variables for R2 cluster was notably lower with AUC=0.66 for age and AUC=0.50 for gender.
A detailed comparison of manual and automatic segmentation for the task of discriminating cluster R2 from all other clusters with respect to sensitivity, specificity, positive predictive value, and negative predictive value is given in Online Resource 1.

\begin{table}[!ht]
    \centering
    \begin{tabular}{l c c c c} \toprule
    Variable & ~~~R1~~~ & ~~~R2~~~ & ~~~R3~~~ & ~~~R4~~~ \\ \midrule
    Deep learning ASD & 0.26 & 0.73 & 0.53 & 0.47 \\
    Deep learning 1/BEVR & 0.23 & 0.80 & 0.36 & 0.53 \\
    Age & 0.29 & 0.66 & 0.72 & 0.41 \\
    Gender & 0.44 & 0.50 & 0.58 & 0.52 \\ \bottomrule
    \end{tabular}
    \caption{ROC AUCs of shape features and demographic variables for the task of discriminating one RNASeq cluster from all others.}
    \label{tab:tab3}
\end{table}

\begin{figure}[!ht]
    \centering
    \includegraphics[width=0.67\linewidth]{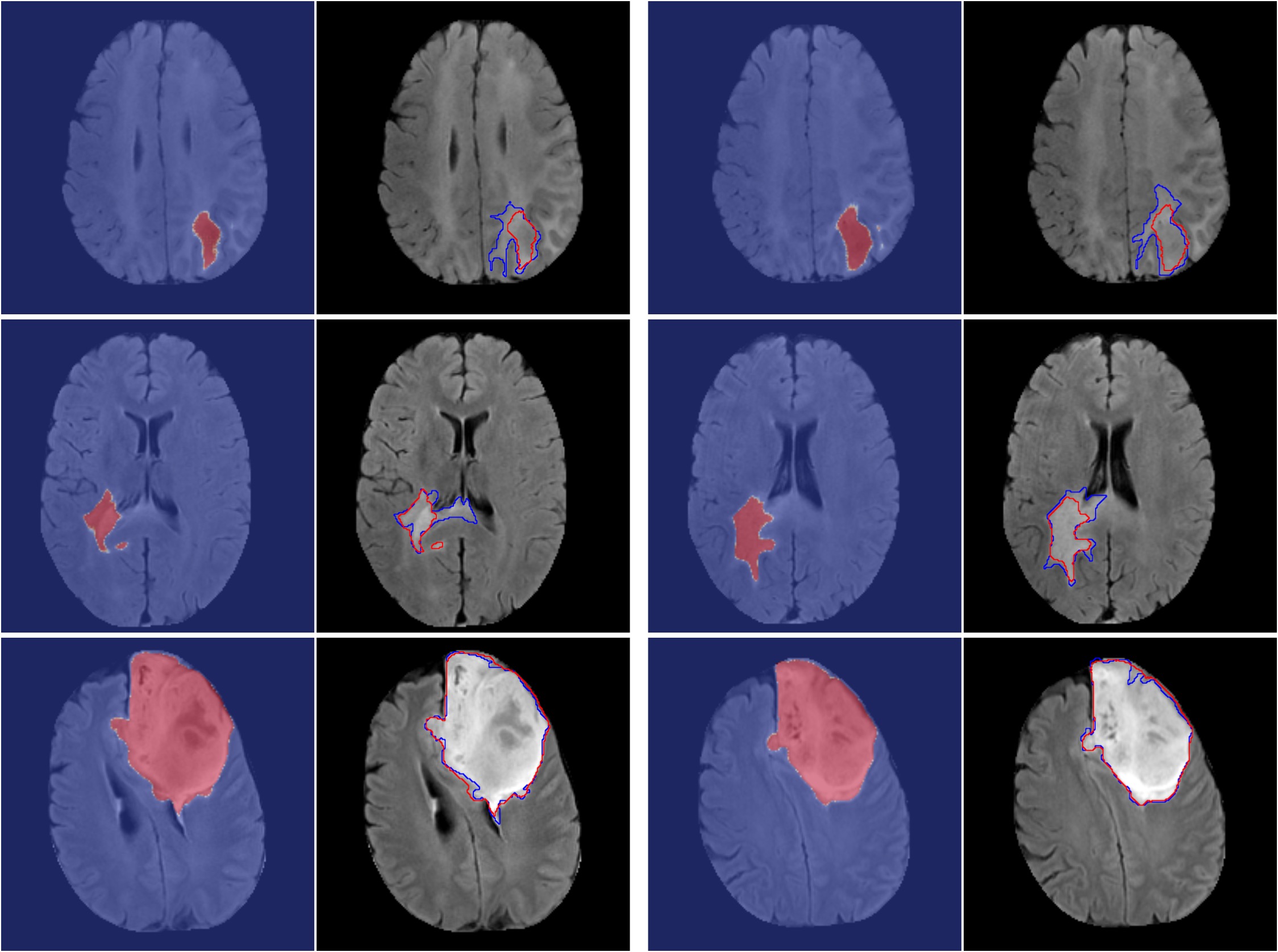}
    \caption{Examples of automatic segmentation for low (top), moderate (middle), and high (bottom) agreement with ground truth. Their Dice coefficient is 50\%, 82\% and 95\% respectively. In each pair, the first image shows a heatmap of raw model output and in the second image blue outline corresponds to ground truth and red to postprocessed automatic segmentation output. Images show FLAIR modality after preprocessing and skull stripping.}
    \label{fig:fig4}
\end{figure}

In terms of tumor segmentation, our deep learning algorithm achieved mean Dice coefficient of 82\% and median Dice coefficient of 85\%.
For the 101 cases with all sequences available, mean and median Dice coefficient was the same as for all 110 cases.
For the remaining 9 cases segmented based on FLAIR sequence only, mean and median Dice coefficient was 82\% and 88\%, respectively.
Examples of segmentations that we obtained alongside with ground truth masks for cases of varying performance of our segmentation algorithm are presented in Figure~\ref{fig:fig4}.
Due to max pooling layers included in the U-Net architecture which allow for processing of large volumes on currently available computers, the automated segmentation is less sensitive to high curvature and sulcation and therefore, the produced masks tend to be more smooth comparing to manual segmentations.

Given small sample size, we additionally assessed the stability and performance of our segmentation model using predictions generated for noisy input. For each input image, we applied additive Gaussian noise with zero mean and standard deviation equal to 10\% and 20\% of standard deviation computed on the training data.
In the first case with 10\% noise level, mean Dice coefficient decreased to 81\%.
After increasing noise level to 20\%, mean Dice coefficient further decreased to 79\%.
In both cases, median Dice coefficient was 85\%.

\section{Discussion}
\label{sec:discussion}

In this study, we were able to demonstrate that fully automatically-assessed imaging features of low grade gliomas are associated with tumor molecular subtypes established using genomic analysis.
The strength of these associations was shown to be moderate. Deep learning algorithms were used to segment the tumors.

Using imaging to predict tumor genomics is of very high importance and if accurate models are developed, it could be incorporated in the current treatment paradigm in a variety of ways.
In the simplest scenario, if the model is highly accurate, it could simply replace the genomic analysis altogether.
Current state of the art radiogenomic models, such as the one shown in this paper which represents moderate predictive performance, do not yet justify such substitution.
However, there are other ways in which even models with moderate performance could contribute valuable information.
The imaging data is available early in the process and therefore approximate assessment of tumor biology before the surgery could still be of help in guiding the next steps.
The approximate imaging surrogate of molecular subtype would also be of particular use for patients that do not immediately undergo surgical excision of the tumor.
In such case, in the absence of tissue analysis, the approximate classification by imaging could be of very high value since genomic subtypes are highly correlated with patient outcomes.
If biopsy results are available for a tumor that has not been fully resected, the imaging surrogate of subtype can still be of use given a potential high intra-tumor heterogeneity of the lesions and therefore a possibility that a local biopsy does not accurately reflect the overall genomics of a tumor.
Imaging offers a complete view of a tumor.
Finally, even if the overall accuracy is not perfect, it might be possible to operate at a high positive predictive value or high predictive value and the surrogate imaging-based models of genomic subtypes could be still used for triaging patients for genomic tests, even if it is a small minority of the patients.
For example, if based on imaging there is a high confidence that a tumor is of a particular aggressive subtype, the patient could be treated accordingly without additional expensive and invasive genomic testing.
If on the other hand the imaging-based marker has low confidence, then additional genomic test could be ordered.

An important step toward accurate and reproducible assessment of imaging features of lower-grade gliomas is accurate segmentation of the tumors.
While the annotation by radiologists is considered a gold standard, a considerable inter observer variability has been documented for this task.
For the whole tumor segmentation of LGG on brain MRI, Dice coefficient between two expert raters is 84\% with standard deviation of 2\%~\cite{menze2014multimodal}.
This demonstrates that our algorithm falls within acceptable level precision. At the same time, our algorithm provides a fully reproducible and consistent way of tumor quantification for future cases.

Automatic segmentation of tumors such as the one showed in this study has multiple advantages.
First, it addresses the inter-observer variability described above.
Since there is only one reader (the computer algorithm), the inter-observer variability is non-existent.
Furthermore, it addresses the problem of intra-observer variability.
The algorithm is deterministic which means that given the same image, the algorithm will always perform an identical assessment.
Finally, application of a computer algorithm is inexpensive and fast.
The performance of our segmentation algorithm in terms of the mean Dice coefficient was 82\% which puts it on a par with expert human readers. 
This was achieved with the help of deep learning which has demonstrated similar phenomenal performance in other applications. 

Our results show that RNASeq R2 cluster, compared to other clusters, is associated with the tumors of notably higher irregularity of shape as quantified by bounding ellipsoid volume ratio, angular standard deviation and margin fluctuation.
R2 cluster is in turn linked to considerably poorer overall survival as compared with R1, R3, and R4~\cite{cancer2015comprehensive}.
The same conclusion can be drawn for molecular subtype of IDH wild type which indicates less favorable prognosis that is close to glioblastoma prognosis~\cite{cancer2015comprehensive}.
It is associated with relatively high angular standard deviation. This is consistent with findings obtained in the previous study for shape features extracted from manually segmented tumors~\cite{mazurowski2017radiogenomics}.
This points out to a conclusion that angular standard deviation, margin fluctuation, bounding ellipsoid volume ratio, and potentially other features that measure the irregularity of tumor shape may be prognostic of patient’s outcome.

\section{Limitations}
\label{sec:limitations}

This study had limitations.
It constitutes only a first step toward imaging-based surrogates of genomic subtypes.
Specifically, only three imaging features were considered.
While these features were selected based on prior evidence of their effectiveness and therefore are of high importance, they constitute a small sample of different features that could be calculated including texture and enhancement of the tumor and its surroundings.
Furthermore, a fairly limited sample size was used in the study (110 patients) since data that contains comprehensive genome-wide assays alongside with imaging is still rare.
While no separate validation set was available in this study, we utilized a commonly used cross-validation technique, which splits the data into training and test sets to avoid a positive evaluation bias. 

Regarding segmentation algorithms, there are many methods for performing automatic segmentation of brain tumors that could be considered for comparison and to further improve our results~\cite{litjens2017survey, akkus2017deep}.
First, regarding general approaches to segmentation, in~\cite{havaei2017brain} deep learning with sliding-window approach was used.
An improvement that significantly reduces computational complexity is a fully convolutional neural network which allows for processing entire image in one forward pass~\cite{badrinarayanan2017segnet, long2015fully}.
In U-Net architecture, used in our study, additional skip connections between encoder and decoder parts of the network are used~\cite{ronneberger2015u}.
Second, regarding network architecture, different types have been proposed, e.g. ResNet~\cite{zhang2017image}, Inception~\cite{szegedy2016rethinking}, and DenseNet~\cite{jegou2017one}, which were incorporated in segmentation models.
Finally, various optimization functions for training deep learning segmentation models were proposed.
The most commonly applied is cross-entropy loss, used also in classification models.
However, for highly imbalanced segmentation tasks, loss functions based on Dice similarity coefficient outperformed other loss functions in many applications~\cite{drozdzal2016importance, sudre2017generalised, fidon2017generalised, zhang2017brain}.

\section{Conclusions}
\label{sec:conclusions}

In conclusion, we demonstrated that features of MRI, extracted in a fully automatic manner using deep learning algorithms, were associated with tumor molecular subtypes of lower-grade gliomas determined using genomic assays.
This shows promise for reproducible non-invasive imaging-based surrogates of tumor genomics in brain cancer.

\vspace{2em}

\bibliography{references}

\end{document}